\documentclass[prb,aps,twocolumn,superscriptaddress,floatfix]{revtex4-2}
\usepackage[english]{babel}
\usepackage{amsmath} 
\usepackage{amssymb}
\usepackage{graphicx}
\usepackage{dcolumn}
\usepackage{bm}
\usepackage[colorlinks=true, linkcolor=blue, citecolor=blue, urlcolor=blue]{hyperref}
\usepackage{overpic}

\begin{document}

\title{Synchronization and Spatiotemporal Phases in Driven Precessing Domain Walls}

\author{Luz P\'erez Herrera}
\affiliation{Instituto Balseiro, UNCUYO, (R8402AGP) San Carlos de Bariloche, R\'{\i}o Negro, Argentina}

\author{E. F. Pusiol}
\affiliation{Centro At\'omico Bariloche, CNEA-CONICET, (R8402AGP) San Carlos de Bariloche, R\'{\i}o Negro, Argentina}

\author{A. B. Kolton}
\affiliation{Centro At\'omico Bariloche, CNEA-CONICET, (R8402AGP) San Carlos de Bariloche, R\'{\i}o Negro, Argentina}
\affiliation{Instituto Balseiro, UNCUYO, (R8402AGP) San Carlos de Bariloche, R\'{\i}o Negro, Argentina}

\date{\today}

\begin{abstract}
The dynamics of extended magnetic domain walls driven by constant and alternating magnetic fields are studied using a collective coordinate $u$--$\phi$ model. While purely constant-field driven systems exhibit deterministic spatiotemporal chaos and roughening, the alternating component can lock the precessional frequency via Shapiro-like steps. Using linear stability analysis and Floquet theory, the phase boundaries of stable, synchronized flat wall states are mapped. By studying the nonlinear dynamics of the extended system, a rich phase diagram is identified, composed of five dynamical regimes characterized by the combination of domain wall spatial order (flat, rough, or periodic) and domain wall precession temporal order (pinned, synchronized, or unsynchronized). Notably, the macroscopic temporal order of the synchronized phases can lead to a flat domain wall or to a periodic structure dominated by a narrow band of linearly unstable modes, both with coherent precession of the domain wall magnetization, while unsynchronized phases can be flat, with coherent precession, or rough with spatiotemporal chaos and incoherent precession.
\end{abstract}

\maketitle

\section{Introduction}

Synchronization is a generic phenomenon in nonlinear dynamical systems whereby oscillators adjust their rhythms through mutual interactions or external periodic forcing. It occurs in a wide variety of physical, biological and technological systems, including Josephson junctions, coupled pendulums and neuronal networks~\cite{Strogatz2015,Pikovsky_Rosenblum_Kurths_2001}.

Magnetic domain walls (DWs) can also behave like oscillators. When driven by a field higher than the Walker breakdown field~\cite{Schryer1974}, the internal magnetization angle begins to rotate continuously at a specific angular velocity. Within the framework of the 1D Walker model, adding a harmonic component to the driving field produces Shapiro steps~\cite{shapiro1963josephson,berger1991prediction,kim2019}: the precessional frequency of the internal degree of freedom remains fixed to an integer multiple of the driving frequency over a range of parameters.

While insightful, the 1D Walker model fails to capture the effects of the DW flexural modes~\cite{thevenard2011domain,gourdon2013domain,Kim2024}. In a more realistic model for an extended DW, it has been recently shown~\cite{pusiol2026deterministic} that spatiotemporal chaos can emerge when the wall enters the precessional regime. Even in this complex condition, the corresponding rigid wall frequency still appears in the power spectrum above the chaotic broadband baseline noise.

This naturally raises the question of whether synchrony can occur in extended DWs. In that case, the macroscopic order induced by the synchrony should manifest itself in the spatial structure of the DW. Similar effects have been found in simulations~\cite{koltonmodelocking,koltonmodelocking2} and experimentally~\cite{okumasuppresion} in other physical systems susceptible to synchrony. An alternating magnetic field or electrical current via spin transfer torque could thus serve to control the otherwise complex DW dynamics, a capability highly relevant for DW-based technologies~\cite{Parkin2008,kim2019,Moon_2017,Munsu2021}.

In this work, we study the dynamics of extended DWs with perpendicular magnetic anisotropy using the collective-coordinate $u-\phi$ model~\cite{Slonczewski1972, Schryer1974}. We analyze the effects of an AC driving field on the spatiotemporal dynamics of the DW over a wide range of system sizes. In particular, we demonstrate that an appropriate AC field can suppress the spatiotemporal chaos of the DC-driven wall. We identify the parameter regions in which the DW synchronizes with the external drive and determine how synchronization affects its long-range spatial structure. This reveals a phase diagram with five distinct dynamical regimes, characterized by the combination of the DW spatial structure (flat, rough or periodic) and the temporal dynamics of the DW precession (pinned, synchronized or unsynchronized). In the thermodynamic limit, these regimes are separated by sharp boundaries, which we expect to correspond to nonequilibrium dynamical phase transitions.

\section{Models and Methods}
Our fundamental description of the DW dynamics is based on the Landau-Lifshitz-Gilbert (LLG) equation:
\begin{equation}
\dfrac{\partial\mathbf{m}}{\partial t}= -\gamma\mathbf{m}\times\mathbf{H}_{\text{eff}}+\alpha\mathbf{m}\times\dfrac{\partial\mathbf{m}}{\partial t} \, ,
\label{eq_LLG}
\end{equation}
where $\gamma$ is the reduced gyromagnetic ratio and $\alpha$ is the Gilbert damping constant. The unit magnetization vector is defined in spherical coordinates as $\mathbf{m} = (\sin\theta \cos\varphi, \sin\theta \sin\varphi, \cos\theta)$. The effective field $\mathbf{H}_{\text{eff}}=-(1/\mu_0M_s) \delta E/\delta \mathbf{m}$ is derived from the total magnetic energy functional, which incorporates exchange, anisotropy, Zeeman, and magnetostatic contributions:
\begin{equation}
\begin{split}
    \label{eq:energy}
    E =& \int \Bigl\{  A_\text{ex}[(\nabla\theta)^2+\sin^{2}\theta(\nabla\varphi)^2]-K_u\cos^2\theta  \\
    -&\mu_0H_z(t)M_s\cos\theta +\dfrac{1}{2}\mu_0M_s^{2} [N_{n}(\mathbf{m} \cdot \mathbf{n})^{2}+\cos^{2}\theta] \Bigl\}d\mathbf{r} \, .
\end{split}
\end{equation}
Here $M_s$ is the saturation magnetization, $A_\text{ex}$ is the exchange stiffness constant, $K_u$ the uniaxial anisotropy constant and $\mu_0$ the magnetic permeability. We consider a uniform, time-dependent magnetic field $H_z(t)$ along the $z$-axis. The last contribution in Eq.~\eqref{eq:energy} accounts for magnetostatic effects of a DW in the local thin-film approximation, where $N_n$ is the demagnetizing factor along the DW normal direction~\cite{skaugen2019}. We assume a directed DW with its average normal in the direction ${\mathbf n}\equiv \hat{\mathbf{y}}$.

To gain additional insight, we describe the DW dynamics using a reduced collective-coordinate framework known as the $u-\phi$ model. This approach captures the most relevant degrees of freedom while remaining computationally efficient compared to full micromagnetic simulations of the LLG equation.

\begin{figure}[tb]
    \centering
    \includegraphics[width=\linewidth]{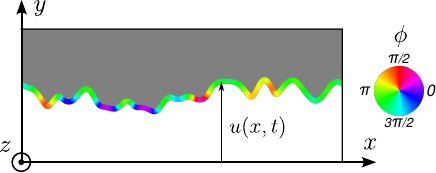}
    \caption{Snapshot of a DW configuration highlighting its collective coordinates $u(x,t)$ and $\phi(x,t)$. Reproduced from Ref.~\onlinecite{pusiol2026deterministic}.}
\label{fig:snap}
\end{figure}

Substituting a Bloch domain-wall ansatz,
\begin{align}
    \theta(x, y, t) &= 2 \arctan\left\{ \exp[{(y - u(x, t))}/{\Delta}] \right\},     \label{eq:Bloch1} \\
    \label{eq:Bloch2}
    \varphi(x, y, t) &= \phi(x, t), 
\end{align}
into Eq.~\eqref{eq_LLG} gives the reduced coordinate model for the DW. Here $u(x, t)$ and $\phi(x, t)$ represent the local position and the internal magnetization angle of the wall, respectively (see Fig.~\ref{fig:snap}). We assume a constant DW width $\Delta = \sqrt{A_\text{ex}/\left(K_u - \mu_0 M_\text{s}^2/2\right)}$, where the denominator accounts for the effective anisotropy including the magnetostatic demagnetizing energy. 

This model treats the DW as a single-valued directed interface, thereby prohibiting overhangs and loops while allowing for the proliferation and dynamics of vertical Bloch lines. It therefore goes beyond standard elastic interface models by incorporating an internal degree of freedom. We do not consider here the additional effects of thermal fluctuations and quenched disorder.

To simplify the analysis, we nondimensionalize the model using characteristic length, time, and magnetic-field scales~\cite{pusiol2026deterministic}. Distances are measured in units of $u_0 = \Delta$ along the $y$-direction, and $L_0 = \Delta ({2\tilde{K}/N_n})^{1/2}$ in the $x$-direction, where $\tilde{K} = 2 K_u / (\mu_0 M_{s}^2) - 1$. Time is measured in units of $T_0 =(1+\alpha^{2})/(\gamma M_s N_n)$, and the applied field $H_{z}(t)$ is expressed in units of the Walker breakdown field~\cite{skaugen2019} $H_W = \alpha M_s N_n / 2$, with $h(t)=H_{z}(t)/H_{W}$. See Appendix~\ref{ap:exp_cond} for details on the conversion to physical units for a thin Co/Pt multilayer. The resulting equations constitute the $u$--$\phi$ or Slonczewski model~\cite{Slonczewski1972}:
\begin{align}
\dot{u} &= \frac{1}{2} \left[ \alpha^2 h(t) + \sin(2\phi) + \alpha \partial_x^2 u - \partial_x^2 \phi \right], \nonumber \\
\dot{\phi} &= \frac{1}{2} \left[ \alpha h(t) - \alpha \sin(2\phi) + \partial_x^2 u + \alpha \partial_x^2 \phi \right].
\label{eq:uphi}
\end{align}

This model reduces to the Walker model~\cite{Schryer1974} for rigid DWs, where $\partial ^2_xu=\partial^2_x\phi=0$. To study synchronization phenomena in extended DWs we set $h(t)=h_\text{0}+h_\text{a}\cos\left(\Omega t\right)$, with $h_{0}$ the DC field and $h_{\text{a}}$ the AC field amplitude. Note that the unit of frequency in this system is $1/T_0$. We then numerically solve the full system of coupled, non-autonomous nonlinear partial differential equations using a pseudo-spectral Crank--Nicolson scheme (see Appendix~\ref{ap:num_met} for details).

\section{Synchrony in rigid domain walls}
\label{sec:flat_sync}

Given that the extended DW can become flat under certain conditions, it is relevant to study synchrony in the resulting rigid DW, described by the Walker model. This rigid-DW regime represents the simplest form of macroscopic order the DW can exhibit, and is considerably more analytically tractable than the full extended model.

To simplify the analysis, we define the variables $\tau\equiv\Omega t$, $q(x,\tau)\equiv  u(x,\tau/\Omega)$ and $\psi(x,\tau)\equiv2\phi(x,\tau/\Omega)$. If we assume a spatially uniform DW given by $q(x,\tau)=q_1(\tau)$ and $\psi(x,\tau)=\psi_1(\tau)$, Eqs.~\eqref{eq:uphi} yield the following system of coupled ordinary differential equations for $q_1$ and $\psi_1$:
\begin{align}
    \dot{q}_1 &= \frac{1}{2\Omega} \left[ \alpha ^2h_0+\alpha^2h_\text{a}\cos(\tau) + \sin (\psi_1)\right], 
    \label{eq:walker_tau1}
    \\
    \dot \psi_1 &= \frac{\alpha}{\Omega} \left[ h_0 +  h_\text{a}\cos(\tau) - \sin(\psi_1) \right].
    \label{eq:walker_tau2}
\end{align}
The period of the non-autonomous term is fixed to $2\pi$, and Eq.~\eqref{eq:walker_tau2} depends only on the parameters $h_0$, $h_\text{a}$ and the ratio $\alpha/\Omega$. We will focus on this equation since it is decoupled from the $q_1$ solution and dictates the overall dynamics of the system.

For the case $h_\text{a}=0$, Eq.~\eqref{eq:walker_tau2} reduces to the Adler equation~\cite{adler1946study}, for which there is an analytical expression for the time average $\langle\dot\psi_1\rangle_\tau$:
\begin{equation}
    \langle\dot \psi_1 \rangle_\tau  = \begin{cases} 
0 & \text{if } \left| h_0\right| \leq 1, \\ 
\operatorname{sgn}(h_0)\dfrac{\alpha}{\Omega} \sqrt{h_0^2 - 1} & \text{if }\left| h_0\right| >1.
\end{cases}
    \label{eq:avg_walker}
\end{equation}
In addition, the average velocity of $q_{1}$ is given by
\begin{equation}
    \langle \dot{q_{1}} \rangle_{\tau} = \frac{\alpha^{2}+1}{2\Omega}h_{0} - \frac{1}{2\alpha}\langle\dot \psi_1\rangle_{\tau}.
    \label{eq:dot_q1_avg}
\end{equation}
Note that this expression is also true for the center of mass of the flexible wall model with periodic boundary conditions. Therefore, for the time-independent driving case we obtain
\begin{equation}
    \langle\dot q_1 \rangle_\tau  =
    \begin{cases} 
     \dfrac{\alpha^{2}+1}{2\Omega}h_{0} & \text{if } \left| h_0\right| \leq 1 \ , \\ \\
    \dfrac{\alpha^{2}+1}{2\Omega}h_{0}-\dfrac{1}{2\Omega}\operatorname{sgn}(h_0) \sqrt{h_0^2 - 1} & \text{if }\left| h_0\right| >1 \ .
    \end{cases}
    \label{eq:avg_q1_cases}
\end{equation}

When $h_\text{a}\neq0$, solutions must be found numerically (see Appendix~\ref{ap:num_met}). To determine if a stable periodic solution exists, we recorded the $\psi_1$ variable modulo $2\pi$ at every period of the drive: the system has converged to a stable periodic solution if and only if these values converge to a single equilibrium value $\psi_\text{eq}$, independent of the initial condition.

\begin{figure}[tb]
    \centering
    \begin{overpic}[width=\linewidth]{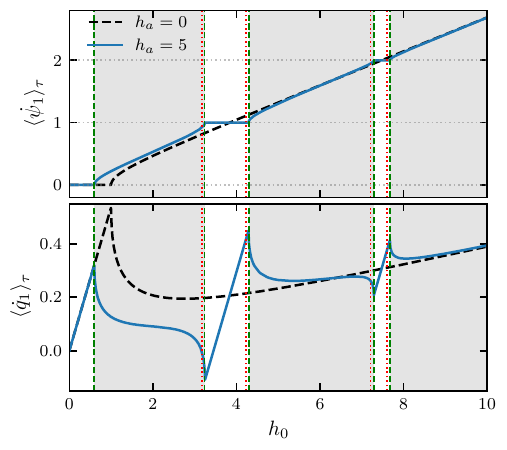}
        \put(89, 52.5){\textbf{(a)}}
        \put(89, 13.5){\textbf{(b)}}
    \end{overpic}
    \caption{
    Steady-state average velocities of the (a) internal ($\psi_1$) and (b) translational ($q_1$) degrees of freedom as a function of the DC field $h_0$, for $h_\text{a}=0$ (dashed black) and $h_\text{a}=5$ (solid blue). White regions indicate the synchronization steps, where $\langle\dot\psi_1\rangle_\tau$ locks to integer multiples of the driving frequency [dotted horizontal lines in (a)]. Vertical red dotted lines indicate synchrony boundaries as given by Eq.~\eqref{eq:width_steps}; their second-order approximation, given by Eq.~\eqref{eq:h0_2order}, is shown as green dashed lines. Walker-breakdown-like spikes in $\langle\dot q_1\rangle_\tau$ recur in (b) at each unlocking transition of $\psi_1$. 
    }
    \label{fig:ej_walker}
\end{figure}

Since $\psi=2\phi$, locking of $\psi$ to $\langle\dot\psi_1\rangle_\tau=n$ corresponds to a precession of the physical angle $\phi$ at $\langle\dot\phi\rangle=n\Omega/2$. Synchrony is expected to occur near the values of $h_0$ for which Eq.~\eqref{eq:avg_walker} yields an integer. Figure~\ref{fig:ej_walker} shows example curves of the steady-state average velocities of both $q_{1}$ and $\psi_{1}$ as a function of the DC field $h_0$, with the white regions marking the synchronization steps where $\langle \dot{\psi_{1}} \rangle_{\tau} = n$, and $n =0,1,2$. A Walker-breakdown-like feature repeats in $\langle \dot q_1 \rangle_\tau$ at each frequency-unlocking transition of the internal degree of freedom $\psi_1$. Within each synchronization step, the translational average velocity is
\begin{equation}
    \langle\dot q_1 \rangle_\tau  = \dfrac{\alpha^{2}+1}{2\Omega}h_{0}-\dfrac{n}{2\alpha}, \quad h^{n}_{1}\leq h_{0} \leq h^{n}_{2},
\end{equation}
with $h^{n}_{1}<h^{n}_{2}$ delimiting the extension of the $n$-th step. Consequently, the DW mobility within a synchronized state is identical to that of the constant-driving-field case below the Walker field ($|h_0|<1$).

These synchronization regions in $\langle \dot \psi_1 \rangle_\tau$ are known as \textit{Shapiro steps}, as they are analogous to those predicted and observed in the I--V curves of driven overdamped Josephson junctions~\cite{shapiro1963josephson}.

In a synchronized state, solutions should be of the form
\begin{equation}
    \psi_\text{s}(\tau) = \psi_\text{c} + n\tau  + \tilde \psi(\tau),
    \label{eq:theta_sync}
\end{equation}
with $\psi_\text{c}$ a constant, $\tilde\psi(\tau)$ a $2\pi$-periodic function with zero mean, and $n$ an integer. In this case, it is clear that $\langle \dot\psi_\text{s} \rangle_\tau=n$.

To approximate the boundaries of the parameter regions for which synchronized solutions exist, we can treat the nonlinear $\sin\psi_1$ term as a perturbation by defining the scaled variables $\tilde h_0 \equiv \epsilon h_0$, $\tilde h_\text{a} \equiv \epsilon h_\text{a}$, and introducing $\epsilon \equiv \alpha /\Omega$ as an explicit small parameter. Equation~\eqref{eq:walker_tau2} then reads:
\begin{equation}
    \dot{\psi}_1 = \tilde h_0 + \tilde h_\text{a}\cos(\tau) - \epsilon \sin(\psi_1).
    \label{eq:walker_pert}
\end{equation}
We now expand both the solution $\psi_1$ and the DC driving parameter $\tilde h_0$ in power series of $\epsilon$:
\begin{align}
    \psi_1(\tau) &= \psi^{(0)}(\tau) + \epsilon \psi^{(1)}(\tau) + \epsilon^2 \psi^{(2)}(\tau) + \dots, \\
    \tilde h_0 &= \tilde h_0^{(0)} + \epsilon \tilde h_0^{(1)} + \epsilon^2 \tilde h_0^{(2)} + \dots
    \label{eq:expansion2}
\end{align}
Substituting these expansions into Eq.~\eqref{eq:walker_pert} and collecting terms of order $\epsilon^0$, we obtain the unperturbed equation:
\begin{equation}
    \dot{\psi}^{(0)} = \tilde h_0^{(0)} + \tilde h_\text{a} \cos(\tau).
\end{equation}

Phase-locking to the $n$-th step requires $\langle \dot{\psi}^{(0)} \rangle_\tau = n$, which fixes $\tilde h_0^{(0)} = n$. Integrating yields the zeroth-order solution:
\begin{equation}
    \psi^{(0)}(\tau) = n\tau + \tilde h_\text{a} \sin(\tau) + \psi_\text{c}.
    \label{eq:walker_zeroth}
\end{equation}

At order $\epsilon^1$, the dynamics of the first-order correction $\psi^{(1)}$ are governed by:
\begin{equation}
    \dot{\psi}^{(1)} = \tilde h_0^{(1)} - \sin\left(\psi^{(0)}\right).
    \label{eq:dot_psi_1}
\end{equation}
For the synchronized solution to remain valid at this order, $\psi^{(1)}$ must remain a $2\pi$-periodic function of zero mean. Therefore, the time average of the right-hand side of Eq.~\eqref{eq:dot_psi_1} must vanish:
\begin{equation}
    \tilde h_0^{(1)} = \left\langle \sin\left(\psi^{(0)}\right) \right\rangle_\tau.
\end{equation}
Expanding $\sin\left(\psi^{(0)}\right)$ using the Jacobi--Anger identity,
\begin{equation}
    \sin\left(\psi^{(0)}\right) = \sum_{m=-\infty}^{\infty}J_{m} (\tilde h_\text{a})\sin[(n+m)\tau+\psi_\text{c}],
\end{equation}
we see that every term oscillates with zero mean except the one with $m=-n$. This DC component yields:
\begin{equation}
    \tilde h_0^{(1)} = J_{-n}(\tilde h_\text{a})\sin(\psi_\text{c}) = (-1)^{n}J_{n} (\tilde h_\text{a})\sin(\psi_\text{c}).
    \label{eq:firstorder}
\end{equation}
Truncating the expansion~\eqref{eq:expansion2} at first order gives $\tilde h_0 \approx n + \epsilon \tilde h_0^{(1)}$. Substituting back the original physical parameters $h_0, h_\text{a}$, $\alpha/\Omega$, and the solution~\eqref{eq:firstorder} for $\tilde h_0^{(1)}$, we obtain:
\begin{equation}
    h_{0} \approx \frac{n\Omega}{\alpha} +(-1)^{n}J_{n} \left(\frac{\alpha}{\Omega}h_\text{a}\right)\sin(\psi_\text{c}). 
\end{equation}
Since $\sin \psi_\text{c}\in [-1,1]$, a synchronized solution exists only if
\begin{equation}
    \left| h_{0} - \frac{n\Omega}{\alpha} \right| \leq \left|J_{n}\left(\frac{\alpha}{\Omega}h_\text{a}\right) \right|.
    \label{eq:walkerwithAC}
\end{equation}
Finally, each step is centered at $h_{0}=n\Omega/\alpha$ and has a field-range width 
\begin{equation}
    h^{n}_{2}-h^{n}_{1}=2 \left|  J_{n}\left(\frac{\alpha}{\Omega}h_\text{a}\right) \right|.
    \label{eq:width_steps}
\end{equation}

The details of the continuation of this perturbative treatment to second order in $\epsilon$ are worked out in Appendix~\ref{ap:segundo_orden}. The only second-order effect is a rigid shift depending only on $\alpha h_\text{a}/\Omega$ in the synchronization steps with $|n|>0$. The correction is more significant for the smaller absolute values of $n$ and $h_\text{a}$, vanishing in the limits where either variable goes to infinity. For the $n=0$ step, the second-order corrections vanish identically. In Fig.~\ref{fig:ej_walker}, vertical red dotted and green dashed lines correspond to first- and second-order approximations, respectively.

\section{Linear stability of flat, synchronized domain walls}
\label{sec:linearstabilitydiagrams}

\begin{figure}[tb]
    \centering
    \begin{overpic}[width=\linewidth]{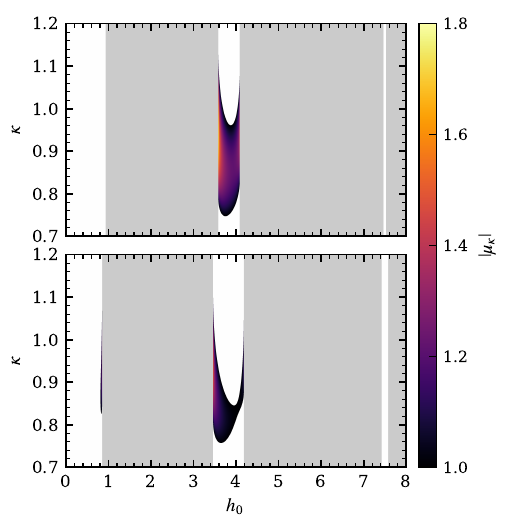}
        \put(14, 90){\textbf{(a)} $h_\mathrm{a}=2$}
        \put(14, 46){\textbf{(b)} $h_\mathrm{a}=3$}
    \end{overpic}
    \caption{
    Linear stability diagram of DW modes for $\alpha=0.27$ and $\Omega = 1$, with $h_\text{a} = 2$ in (a) and $h_\text{a}=3$ in (b). White regions are stable; gray regions do not admit a stable time-periodic solution for the rigid wall. Color indicates the magnitude of the largest Floquet multiplier for unstable regions with $\left|\mu_\kappa\right|>1$.
    }
    \label{fig:sync_stability}
\end{figure}

To study the linear stability of synchronized flat solutions, we adopt the theoretical framework recently detailed by Pusiol et al.~\cite{pusiol2026deterministic} and extend it to account for a time-dependent magnetic field. We linearize a solution of Eq.~\eqref{eq:uphi} $\mathbf{v}(x,\tau)\equiv(u,\phi)^{T}$ around the flat-wall solution,
\begin{equation}
    \mathbf{v}(x,\tau) = \mathbf{v}_{1}(\tau) + \boldsymbol{\eta}(x,\tau), \quad \boldsymbol{\eta}(x,\tau) \equiv \begin{pmatrix} Q(x,\tau) \\ \Phi(x,\tau) \end{pmatrix},
\end{equation}
where $\mathbf{v}_{1}(\tau)\equiv(q_{1}(\tau),\psi_1(\tau)/2)^{T}$. Substituting into Eq.~\eqref{eq:uphi}, the linearized system reads
\begin{equation}
    \partial_\tau \boldsymbol{\eta} = \frac{1}{2\Omega}\,\mathcal{M}(\tau)\,\boldsymbol{\eta},
    \label{eq:linearized_matrix}
\end{equation}
with
\begin{equation}
    \mathcal{M}(\tau) \equiv
    \begin{pmatrix}
    \alpha\,\partial_x^2 & -\partial_x^2+2\cos\psi_1(\tau) \\
    \partial_x^2 & \alpha\,\partial_x^2-2\alpha\cos\psi_1(\tau)
    \end{pmatrix}.
    \label{eq:stability_matrix}
\end{equation}
We perform a Fourier transform in space, so that $\partial_x^2\to-\kappa^2$. Since Eq.~\eqref{eq:linearized_matrix} is linear with $x$-independent coefficients, distinct modes $\kappa$ decouple, yielding
\begin{equation}
    \partial_\tau \boldsymbol{\eta}_\kappa = \frac{1}{2\Omega}\,\mathcal{M}_\kappa(\tau)\,\boldsymbol{\eta}_\kappa,
    \label{eq:linearized_mode}
\end{equation}
with
\begin{equation}
    \mathcal{M}_\kappa(\tau) \equiv -\kappa^2
    \begin{pmatrix}
    \alpha & -1 \\
    1 & \alpha
    \end{pmatrix}
    + 2\cos\psi_1(\tau)
    \begin{pmatrix}
    0 & 1 \\
    0 & -\alpha
    \end{pmatrix},
    \label{eq:stability_matrix_split}
\end{equation}
where the first term is the $\tau$-independent diffusive coupling between $Q_\kappa$ and $\Phi_\kappa$, and the second is the $2\pi$-periodic modulation set by the rigid-wall background solution $\psi_1(\tau)$. The stability of a linear system with periodic coefficients can be determined using Floquet theory. By Floquet's theorem, a general solution of Eq.~\eqref{eq:linearized_mode} can be written as
\begin{equation}
    \boldsymbol{\eta}_{\kappa}(\tau) = P_\kappa(\tau)\,e^{R_\kappa\tau}\,\boldsymbol{\eta}_\kappa(0),
\end{equation}
where $P_\kappa(\tau)$ is a $2\pi$-periodic matrix and $R_\kappa$ is a constant matrix. The monodromy matrix $M_\kappa \equiv P_\kappa(2\pi)\,e^{2\pi R_\kappa}$ is obtained by integrating Eq.~\eqref{eq:linearized_mode} over one period, using the canonical initial conditions $\boldsymbol{\eta}_\kappa(0) = (1,0)^T$ and $(0,1)^T$. The amplitude of the perturbation grows exponentially -- indicating that the system is unstable -- whenever the Floquet multiplier (eigenvalue of $M_\kappa$) of largest modulus, $\mu_\kappa$, satisfies $|\mu_\kappa|>1$. Since the modes $\kappa$ are decoupled, their Floquet multipliers can be computed independently, allowing straightforward parallelization across $\kappa$. Furthermore, the rescaled time conveniently fixes the integration interval to $2\pi$ regardless of parameter values.

Figure~\ref{fig:sync_stability} shows the resulting stability diagrams obtained numerically for $h_\text{a}=2$ (a) and $h_\text{a}=3$ (b), at $\alpha=0.27$ and $\Omega=1$, as an illustrative example. Although the rigid DW is synchronized within the corresponding Shapiro steps, finite intervals of $h_0$ exhibit spatial instabilities. These unstable regions are generally located near the boundaries of the synchronization steps and are characterized by a finite range of wavenumbers with Floquet multipliers satisfying $|\mu_\kappa|>1$. Thus, synchronization of the internal precessional dynamics of the rigid DW does not necessarily imply spatial stability of the extended flexible DW. Instead, a synchronized flat DW can coexist with a band of unstable spatial modes, whose growth rates depend continuously on $\kappa$. The nonlinear evolution of these instabilities, and also the possible instabilities of non-synchronized flat walls (for which Floquet analysis is not adapted), and their effect on the long-time DW structure are investigated in the following section.

\section{Nonlinear dynamics of extended DWs}
\label{sec:nonlinear_dynamics}

We now investigate the long-time dynamics under the combined DC+AC field, going beyond the linear stability analysis of rigid synchronized states. As an illustrative example, we numerically simulate the system of Eqs.~\eqref{eq:uphi} (see Appendix~\ref{ap:num_met}) for the same parameters as in the stability map of Fig.~\ref{fig:sync_stability}(b). We characterize the resulting dynamics through the mean angular velocity $\langle\dot\phi\rangle_{x,t}$, the mean DW velocity $\langle\dot u\rangle_{x,t}$, and the roughness
\begin{equation}
\langle w^2\rangle_t =
\left\langle
\left\langle u^2\right\rangle_x
-\left\langle u\right\rangle_x^2
\right\rangle_t .
\label{eq:roughnessvst}
\end{equation}
The time averages are taken after the system has reached its long-time dynamical regime.

We characterize synchronization with the external drive through the quantity $\langle\sigma^2\rangle_x$, obtained by sampling the DW displacement stroboscopically at intervals of twice the drive period, $T_\text{drive}=2\pi/\Omega$. At each point $x$ we compute the variance over time of $\Delta u(x,t)=u(x,t+2T_\text{drive})-u(x,t)$,
\begin{equation}
  \sigma^2(x)=\overline{\Delta u^{2}}-\overline{\Delta u}^{\,2},
\end{equation}
where the overbar denotes the average over the sampling times $t_m=t_0+2mT_\text{drive}$ ($m=0,1,2,\dots$), and then average along the wall, $\langle\sigma^2\rangle_x=L^{-1}\!\int_0^L\sigma^2(x)\,dx$. When the DW is phase-locked to the drive, $\Delta u(x,t)$ takes the same value at every sampling time, so that $\langle\sigma^2\rangle_x\approx 0$. Otherwise, $\Delta u$ fluctuates from one sample to the next and $\langle\sigma^2\rangle_x$ takes a finite value. Sampling at $2T_\text{drive}$, rather than at $T_\text{drive}$, allows us to detect locking to both the harmonic (period $T_\text{drive}$) and subharmonic (period $2T_\text{drive}$) responses.

To highlight finite-size effects, we consider a small system with $L/L_0=8$, for which the first nonzero mode is $\kappa=2\pi/8\approx 0.785$, and a large system with $L/L_0=1024$. The latter provides access to the extended-wall regime in which deterministic spatiotemporal chaos occurs under purely DC driving~\cite{pusiol2026deterministic}, whereas the small system is strongly affected by the discrete set of available spatial modes. This comparison allows us to distinguish finite-size effects from the behavior associated with a broad spectrum of flexural modes.

\begin{figure}[tb]
\centering
\begin{overpic}[width=\linewidth]{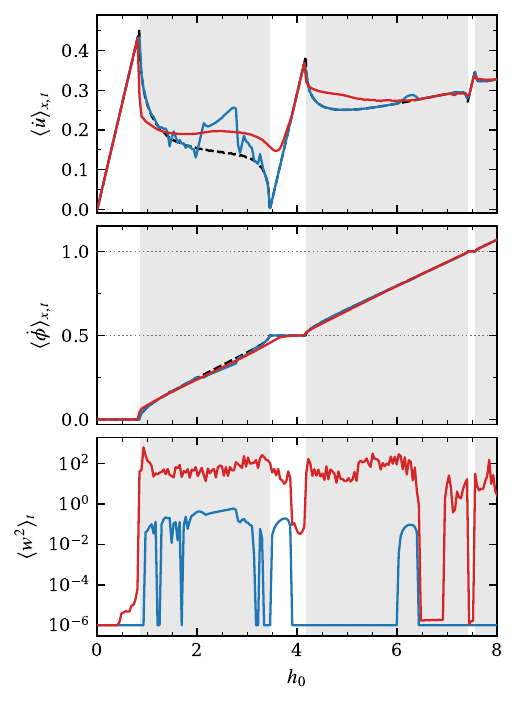}
    \put(15, 94){\textbf{(a)}}
    \put(15, 62){\textbf{(b)}}
    \put(15, 33){\textbf{(c)}}
\end{overpic}
\caption{Flexible extended domain wall dynamics. (a) Mean DW velocity $\langle\dot u\rangle_{x,t}$ and (b) mean precessional velocity $\langle\dot\phi\rangle_{x,t}$ for extended walls of sizes $L/L_0=8$ (blue continuous line) and $1024$ (red continuous line), together with the rigid-wall result (dashed black lines). (c) Steady-state roughness $\langle w^2\rangle_t$ for the two extended-wall sizes ($L/L_0=8$ in blue, $L/L_0=1024$ in red). Values smaller than $10^{-6}$ are shown at $10^{-6}$ for visualization. White and gray regions correspond to synchronized and unsynchronized rigid-wall solutions, respectively. Horizontal dotted lines in (b) indicate the Shapiro steps at $\langle\dot\phi\rangle_{x,t} = \Omega/2 = 0.5$ and $\Omega = 1$. All simulations were performed with $\alpha=0.27$, $\Omega=1$, and $h_\text{a}=3$.}
\label{fig:wall_metrics}
\end{figure}

Figure~\ref{fig:wall_metrics} summarizes the steady-state response of extended domain walls as a function of $h_0$, comparing the large and small systems. The rigid-wall dynamics are also included for comparison (dashed black lines). In Fig.~\ref{fig:wall_metrics}(b), both extended walls display synchronization steps at $\langle\dot\phi\rangle_{x,t}=\Omega/2$ and $\Omega$, although the step is narrower for the large DW. Figure~\ref{fig:wall_metrics}(a) shows that $\langle\dot u \rangle_{x,t}$ increases linearly with $h_0$ in the synchronized regions and displays a Walker-breakdown-like feature at the frequency unlocking edges. In the non-synchronized regions, both for the large and the small systems, $\langle\dot u \rangle_{x,t}$ differs appreciably from the rigid-wall solution. The small system displays a jerky response in these regions associated with a sparse discrete set of unstable modes, while the large system displays a smoother response due to its higher density of unstable modes. These instabilities are reflected in the high values of $\langle w^2 \rangle_{t}$ in the non-synchronized regions in Figure~\ref{fig:wall_metrics}(c). Interestingly however, inside the synchronized regions or Shapiro steps, we can also observe a large departure from the rigid-wall solution for both the small and large systems. This behavior can be linked to the linear Floquet instability shown in Figure~\ref{fig:sync_stability}(b), displaying a band of unstable modes within the Shapiro step, and the differences between the small and large systems can be traced back to the discrete set of modes overlapping such continuous band predicted by Floquet theory. We note that the presence of the band can be innocuous for the small systems if none of its discrete modes overlap the band, as observed in the jump to zero of $\langle w^2 \rangle_{t}$ in the middle of the step, for $L/L_0=8$.

In summary, Figure~\ref{fig:wall_metrics} illustrates different kinds of dynamical regimes which can differ from and sometimes coincide with the rigid-wall solution under combined DC and AC drives. In particular, we observe that the full dynamical properties of the $u-\phi$ model can be related to mode linear instabilities, as for the pure DC case~\cite{pusiol2026deterministic}, within the Shapiro steps. At variance with the DC case, however, Floquet theory cannot be applied outside the steps.

\section{Dynamical Phases}
\label{sec:dynamical_phases}

\begin{figure}[tb]
    \centering
    \includegraphics[width=\linewidth]{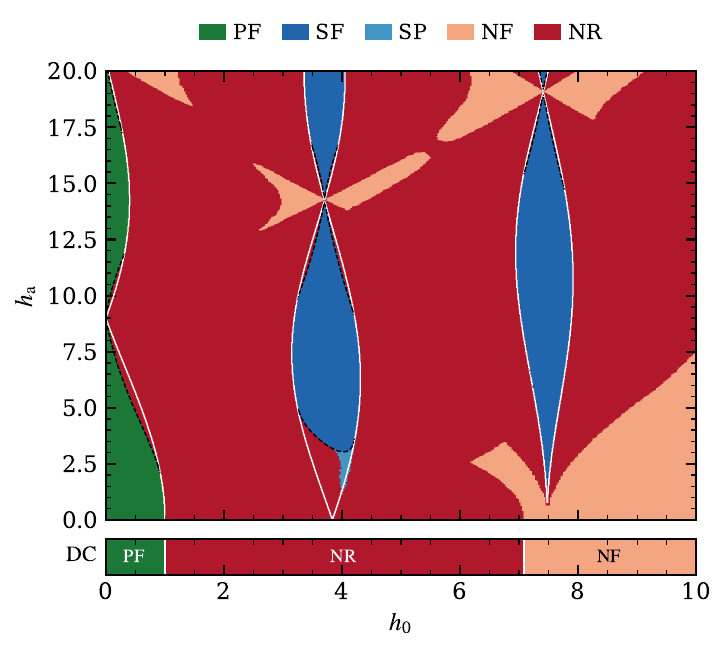}
    \caption{Dynamical phase diagram for an extended domain wall in the large-system limit ($L/L_0 = 256$), showing the five dynamical regimes: pinned flat (PF), synchronized flat (SF), synchronized periodic (SP), non-synchronized flat (NF), and non-synchronized rough (NR), for $\alpha=0.27$ and $\Omega=1$. Continuous white and dashed black curves represent the boundaries of synchrony and linear stability ($\mu=1$) of the corresponding rigid flat wall, respectively. The horizontal strip at the bottom displays the purely DC-driven phases ($h_\mathrm{a}=0$): PF for $0 \le h_0 \le 1$, precessional rough (NR) for $1 < h_0 < h_c$, and precessional flat (NF) for $h_0 \ge h_c$, with $h_c \approx 7.08$. The small-system phase diagram ($L/L_0=8$) is presented separately in Appendix~\ref{ap:finitesizephasediagram}.}
    \label{fig:fases}
\end{figure}

We are now in a position to discuss the main result of this work: the full dynamical phase diagram in the $(h_0,h_\text{a})$ plane at fixed AC frequency. Figure~\ref{fig:fases} shows five distinct dynamical regimes: pinned flat (PF), synchronized flat (SF), synchronized periodic (SP), non-synchronized flat (NF), and non-synchronized rough (NR). The phase diagram corresponds to $L/L_0=256$, for which finite-size effects are negligible on the scale considered. Finite-size effects become visible for substantially smaller systems (see Appendix~\ref{ap:finitesizephasediagram}), so we use the $L/L_0=256$ results as representative of the $L\to\infty$ dynamical phase diagram. 

Analysis of the distribution of $\langle\sigma^2\rangle_x$ reveals it is sharply bimodal, with a six-decade-wide empty gap between $10 ^{-12}$ and $10^{-6}$ containing less than $0.02\%$ of all evaluated points. Similarly, the roughness $\left\langle w^2\right\rangle_t$ distribution displays a gap between $10^{-6}$ and $10^{-4}$ containing only $0.024\%$ of all evaluated points. Consequently, the phase boundaries are only weakly dependent of the threshold choice within the gaps. In this work we chose $\left\langle\sigma^2\right\rangle_x\leq3.5\times10^{-12}$ and $\langle w^2\rangle_x\leq1.0\times10^{-4}$ as the thresholds for synchrony and flatness, respectively. The PF phase was identified as a flat phase with zero net angular velocity $\langle\dot\phi\rangle_{x,t}\leq1.0\times10^{-5}$.

The continuous white and dashed black lines in Fig.~\ref{fig:fases} indicate, respectively, the synchronization boundaries and the linear-stability boundaries of the corresponding synchronized flat rigid-wall solutions. These boundaries provide a reference for the long-time dynamics of the full nonlinear extended-wall equations: the continuous lines delimit the parameter regions where the corresponding rigid-wall solutions are synchronized, while the dashed lines indicate where these synchronized flat solutions become spatially unstable.

In the PF phase, $\langle\dot{\phi}\rangle=0$, so the DW does not exhibit average internal precession, although its mean position $\langle u\rangle$ can still evolve in time. The effective Walker breakdown field in the presence of an AC drive follows $h_W \approx |J_0(\alpha h_\text{a}/\Omega)|$, as predicted by the equality in Eq.~\eqref{eq:walkerwithAC} for the particular $n=0$ case. The remaining four regimes combine temporal synchronization with distinct spatial structures. The SF and SP phases are synchronized, whereas NF and NR are non-synchronized. Spatially, SF and NF remain essentially flat, while SP develops a periodic corrugation and NR exhibits roughening. The SP regime occupies a comparatively narrow region, for relatively low values of $h_\text{a}$, at the bottom of the $n=1$ synchronization region.

The synchronization regions of the extended domain wall remain close to those obtained from the rigid-wall dynamics (delimited by the continuous lines) over most of the phase diagram. These lines are described by Eqs.~\eqref{eq:walkerwithAC} and~\eqref{eq:width_steps}. Deviations due to the flexural modes of the extended DW are most pronounced near the edges of the synchronization steps and in the region where the SP state develops. In these regions, the Floquet stability analysis of the synchronized rigid-wall solution provides a direct indication of the onset of spatial instabilities (indicated by the dashed lines). In Fig.~\ref{fig:fases} we also observe that the dashed lines provide an approximation for the onset of instabilities inside the synchronized phase of the rigid DW. Thus, synchronization and spatial stability are related but distinct properties of the driven wall.

The AC drive can qualitatively modify the dynamics observed under pure DC driving. It can suppress roughening and stabilize a synchronized flat state, or it can synchronize a spatially unstable state, leading to the SP regime. Conversely, non-synchronized flat states can also occur under AC driving. The resulting phase diagram therefore reflects the interplay between temporal synchronization and spatial stability rather than a single synchronization transition. Since the flat phases are indistinguishable from the rigid-wall solutions, in the following we solely characterize the non-flat phases.

\begin{table*}[t]
\caption{Characterization of the dynamical phases in terms of the spatial structure of $u$, synchronization of the domain-wall precession $\phi$ with the external drive, and coherence of the precessional dynamics. The PF phase has no average precession of the DW and it remains flat. The SF and SP phases are synchronized and exhibit coherent domain-wall precession, while the NF phase corresponds to coherent but non-synchronized precession. In the NR phase, the precession becomes incoherent and chaotic, accompanied by Edwards--Wilkinson roughening of $u$ in the thermodynamic limit. In the SP phase, the periodic spatial structure is characterized by a wavelength selected by the narrow band of linearly unstable modes.}
\label{tab:phases}
\begin{ruledtabular}
\begin{tabular}{cccc}
Phase & $u$ spatial structure & $\phi$ synchronization & DW average precession \\
\hline
PF & Flat & Pinned & Absent \\
SF & Flat & Synchronized & Coherent \\
SP & Periodic (Floquet-controlled) & Synchronized & Coherent \\
NF & Flat & Non-synchronized & Coherent \\
NR & Rough (Edwards--Wilkinson) & Non-synchronized & Incoherent / chaotic \\
\end{tabular}
\end{ruledtabular}
\end{table*}

We first characterize the NR regime. To quantify its spatial structure, we analyze the roughening process and consider the \textit{time-dependent} structure factor, starting from an initially perturbed flat DW,
\begin{equation}
    S_\kappa(t) = \left\langle\left|\hat u_\kappa(t)\right|^2\right\rangle ,
    \label{eq:factor_estructura}
\end{equation}
and the \textit{time-dependent} global roughness
\begin{equation}
w^2(t) =
\left\langle u^2(t)\right\rangle_x
-\left\langle u(t)\right\rangle_x^2 ,
\label{eq:roughness}
\end{equation}
where the averages are taken over random realizations of the perturbed initial DW.

\begin{figure}[tb]
    \centering
    \begin{overpic}[width=\linewidth]{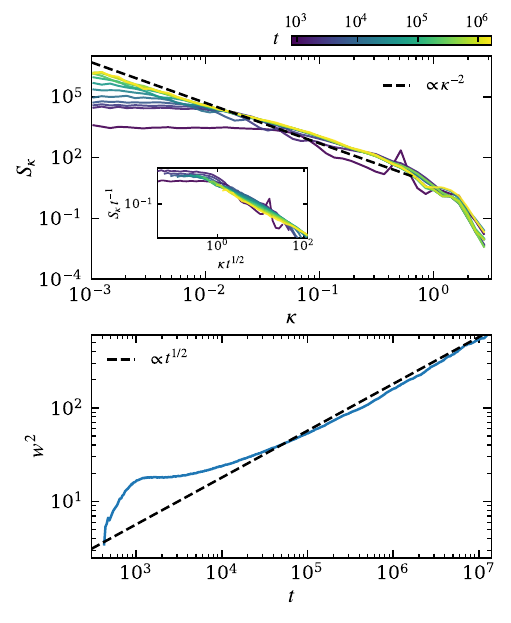}
        \put(72, 58){\textbf{(a)}} 
        \put(72, 13){\textbf{(b)}}
    \end{overpic}
    \caption{Universal scaling and exponent analysis of the non-synchronized rough (NR) phase ($h_0=5,\, h_\text{a} = 3,\, \Omega=1,\, \alpha=0.27$, for $L/L_0=131072$). (a) Main panel: Structure factor $S_\kappa$ as a function of wavenumber $\kappa$ at different times $t$ during the transient growth regime, with the dashed line indicating the asymptotic power-law decay $S_\kappa\propto\kappa^{-2}$. Curves are color-coded by time $t$ according to the colorbar above. Inset: Scaling collapse onto the universal master curve $S_\kappa t^{-1}$ versus $\kappa t^{1/2}$, demonstrating dynamic scaling consistent with $S(\kappa,t)\sim \kappa^{-(1+2\zeta)}F(\kappa t^{1/z})$ with roughness exponent $\zeta=1/2$ and dynamic exponent $z=2$. (b) Domain-wall roughness $w^2$ as a function of time $t$; the dashed line illustrates the characteristic growth regime $w^2\propto t^{1/2}$. These scaling relations are consistent with the NR phase under combined DC+AC driving belonging to the one-dimensional Edwards--Wilkinson universality class.}
    \label{fig:estructura}
\end{figure}

During the transient growth regime that nucleates the NR phase, shown in Fig.~\ref{fig:estructura}, the structure factor follows the scaling form
\begin{equation}
S_\kappa \propto \kappa^{-2}F(\kappa \, l(t)),
\end{equation}
with a growing length scale $l(t)\sim t^{1/2}$, while $F(x)\sim x^2$ for small $x$ and approaches a constant for large $x$. As shown in the inset of Fig.~\ref{fig:estructura}(a), the rescaled data $S_\kappa t^{-1}$ collapse onto a single universal curve when plotted against $\kappa t^{1/2}$. Correspondingly, the roughness grows as
\begin{equation}
w^2 \propto t^{1/2}.
\end{equation}
These scaling properties are consistent with the one-dimensional Edwards--Wilkinson (EW) universality class, where $w^2 \sim t^{2\zeta/z}$ and $S(q)\sim q^{-(1+2\zeta)}F(q t^{1/z})$, with a roughness exponent $\zeta=1/2$ and dynamic exponent $z=2$. The collapse in terms of $\kappa t^{1/2}$ singles out $z=2$ and rules out the Kardar--Parisi--Zhang (KPZ) class, which shares $\zeta=1/2$ in one dimension but has $z=3/2$ and would require the variable $\kappa t^{2/3}$ and $w^2\sim t^{2/3}$. This is expected since, within the $u$--$\phi$ model, tilt invariance ($u\to u+sx$ leaves Eqs.~\eqref{eq:uphi} unchanged, as they depend on $u$ only through $\partial_x^2u$) forbids the KPZ nonlinearity $\lambda(\partial_xu)^2$, irrespective of the internal degree of freedom $\phi$. Geometric corrections beyond the small-slope approximation, which are neglected here, could nevertheless produce a crossover to KPZ at very large scales. The NR phase under DC+AC drive is thus in the same universality class as in the purely DC-driven case~\cite{pusiol2026deterministic}.

\begin{figure}[tb]
    \centering
    \includegraphics[width=\linewidth]{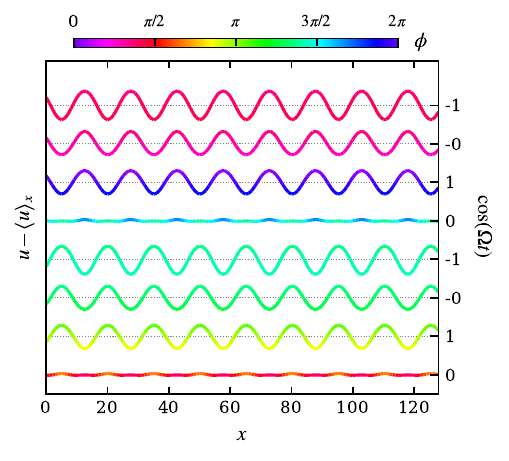}
    \caption{Snapshots of the relative domain-wall displacement $u(x,t)-\langle u\rangle_x$ along a section of the DW in the steady-state synchronized periodic (SP) phase ($h_0=4$, $h_\text{a}=3$, $\alpha=0.27$, and $\Omega=1$), taken at evenly spaced intervals over two full periods of the AC drive (as indicated by the oscillation of $\cos(\Omega t)$ on the right).}
    \label{fig:sc_cascade}
\end{figure}

We next turn to the SP regime. Figure~\ref{fig:sc_cascade} shows representative configurations over one period of the AC drive. The wall develops a periodic corrugation whose amplitude oscillates in time, while the spatial profile recurs periodically. The roughness approaches a finite value as the system size increases, and the dominant wave number remains finite, indicating that the corrugation is not a finite-size artifact.

\begin{figure}[tb]
    \centering
    \begin{overpic}[width=\linewidth]{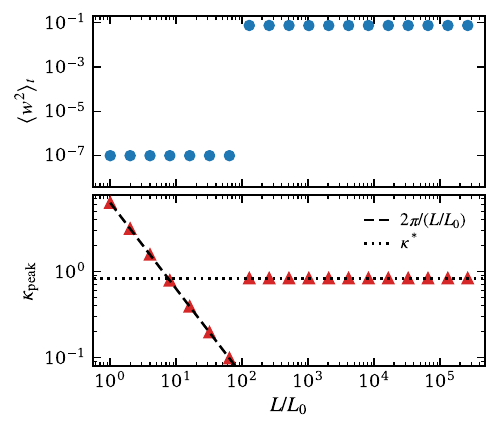}
        \put(90, 52){\textbf{(a)}}
        \put(90, 17){\textbf{(b)}}
    \end{overpic}
    \caption{Finite-size analysis of the synchronized periodic (SP) phase ($h_0=4$, $h_\text{a}=3$). (a) DW steady-state roughness $\langle w^2 \rangle_t$ and (b) wavenumber of the maximum peak in the structure factor $\kappa_\text{peak}$ as a function of $L/L_0$. The long-range spatiotemporal order is dominated by the most linearly unstable mode $\kappa^\ast$ (dotted line) of the synchronized phase. For large enough systems, the roughness converges to a constant value independent of system size. For small sizes, the structure factor is dominated by the lowest available discrete mode $2\pi/(L/L_0)$ (dashed line) and the DW remains flat due to the lack of modes in the narrow instability band.}
    \label{fig:estructuraSCvsL}
\end{figure}

The spatial periodicity of the SP phase is directly related to the Floquet instability of the synchronized flat state. In the unstable region, a finite band of wave numbers has Floquet multipliers larger than unity, with a maximum growth rate at a characteristic wave number $\kappa^*$. The nonlinear SP state is dominated by a wave number $\kappa_{\rm peak}$ that approaches a finite value with increasing system size and is consistent with the Floquet-selected $\kappa^*$ (indicated by the dotted horizontal line). Figure~\ref{fig:estructuraSCvsL} shows that both the roughness and $\kappa_{\rm peak}$ converge as $L$ increases. For sufficiently small systems, the allowed wave numbers may not sample the unstable band adequately; in that case the wall can remain effectively flat even when the corresponding continuum system is spatially unstable. Specifically, at $h_0=4$ and $h_\text{a}=3$, our Floquet calculation yields an unstable band with multiplier $|\mu_\kappa| > 1$ restricted to the narrow range $\kappa \in [0.817, 0.850]$ (bandwidth $\Delta\kappa \approx 0.033$), with a maximum multiplier $|\mu_{\max}| \approx 1.0061$ at $\kappa^* \approx 0.834$. For a small system of size $L/L_0=8$, the discrete allowed wavenumbers are $\kappa_n = 2\pi n / (L/L_0) \approx 0.785 n$. The first mode $\kappa_1 \approx 0.785$ lies below the lower edge of the unstable band by $\approx 0.031$ (with decaying multiplier $|\mu_{\kappa_1}| \approx 0.956 < 1$), while the second mode $\kappa_2 \approx 1.571$ lies far above it. Consequently, no discrete mode falls within the narrow instability band, explaining why the $L/L_0=8$ wall remains strictly flat.

The finite-size analysis therefore demonstrates that the periodic corrugation in the SP regime has an intrinsic spatial scale selected by the linear instability of the synchronized state. In the large-system limit, the SP phase exhibits long-range spatial periodic order together with temporal synchronization.

The main properties of the five dynamical phases are summarized in Table~\ref{tab:phases}.

\section{Conclusions}

We investigated synchronization in extended domain walls driven by combined DC and AC magnetic fields and its relation to spatial order. We showed that synchronization can occur even in very large systems exhibiting spatiotemporal chaos in the purely DC-driven regime. Synchronization is generally accompanied by a suppression of this chaotic dynamics and, in most cases, by a flattening of the domain wall. However, synchronized regimes with a periodically corrugated spatial structure also exist. In this regime, the corrugation is dominated by a finite band of linearly unstable modes with a well-defined maximum growth rate, linking the spatial structure of the nonlinear state to the Floquet instability of the synchronized flat state. The resulting dynamical phase diagram, shown in Figure~\ref{fig:fases}, summarizes these different regimes.

The AC field amplitudes required to drive a chaotic domain wall into a synchronized phase (SF or SP) are of the order of the Walker field. For a thin Co/Pt multilayer, the minimum AC amplitude required to reach the SF phase is of order $10$~mT, with frequencies on the order of $100$~MHz. These conditions are experimentally accessible (see Appendix~\ref{ap:exp_cond} for the correspondence between the model parameters and physical units). The finite-size effects identified in this work may also be relevant for experimentally accessible micrometer-scale systems.

The framework developed here neglects disorder and thermal fluctuations~\cite{ferre2013,ferrero2021creep}, which may affect the experimental observation of the predicted dynamical regimes. Nevertheless, the present theory provides a basis for extending the analysis to more realistic domain-wall models that incorporate these effects. The framework can also be generalized to higher-dimensional geometries and to electrical current drives, providing directions for future studies.

Beyond these extensions, our results demonstrate that a spatially extended nonlinear system exhibiting deterministic spatiotemporal chaos can synchronize to an external periodic drive. We further show that synchronization can suppress the chaotic dynamics while preserving spatial structure, leading either to a synchronized flat state or to a synchronized periodically corrugated state with long-range spatiotemporal periodic order. This coexistence of temporal synchronization and spatial instability provides a distinctive dynamical regime of driven precessing domain walls.

\begin{acknowledgments}
We thank T. Giamarchi for discussions. Simulations used the Gerencia de Física HPC cluster at CAB-CNEA (SNCAD, Argentina).
\end{acknowledgments}

\appendix

\section{Experimental conditions}
\label{ap:exp_cond}

We work out the conversion of our nondimensional model into physical units for the case of a standard Co/Pt multilayer, with parameters~\cite{skaugen2019} $\alpha=0.27$, $A_\text{ex}=1.4 \times 10^{-11}$~J/m, $M_\text{s}=9.1\times 10^5$~A/m, $K_\text{u}=8.4\times 10 ^5$~J/m$^3$ and $N_\text{n}=0.016$.

The length scales for this material are $\Delta\approx6.6$~nm and $L_0\approx8.77\Delta\approx58.0$~nm. This means that representative values of $L/L_0$ such as 8, 256, and 1024 used throughout our work correspond to DW lengths of $\approx 464$~nm, $14.8$~$\mu$m, and $59.4$~$\mu$m, respectively ($L/L_0 = 1000$ corresponds to $58.0$~$\mu$m, while the finite-size scaling in Fig.~\ref{fig:estructuraSCvsL} extends up to $L/L_0 \approx 2.6\times 10^5 \approx 15$~mm). The limiting thermodynamic behavior begins to occur for lengths of order $100\, L_0\approx5.8$~$\mu$m.

The time unit in this system is $T_0 \approx0.33$~ns, which corresponds to an angular-frequency unit $1/T_0\approx3.03\times10^{9}$~s$^{-1}$. The angular frequency $\Omega=1$ used in this work corresponds to a real frequency of 480~MHz. The wavelength corresponding to this frequency is $0.6$~m, many orders of magnitude above both length scales of the system. 

The Walker breakdown field is $H_\text{W}\approx2.48$~mT. As a specific example, to drive the system in the SF regime at $H_0=4H_\text{W}\approx10$~mT, with a frequency of $\Omega=1$ (480~MHz), the AC field amplitude should be at least $H_\text{a}\approx3.5H_\text{W}\approx9$~mT.

\section{Numerical methods}
\label{ap:num_met}

All systems of ordinary differential equations (ODEs) were solved using a fourth-order Runge-Kutta method~\cite{press2007numerical}. This includes the system of two coupled, non-autonomous nonlinear ODEs for the Walker model of rigid DWs, Eqs.~\eqref{eq:walker_tau1} and~\eqref{eq:walker_tau2}, and the linear system of ODEs, Eq.~\eqref{eq:linearized_mode}, used to find the Floquet multipliers of the synchronized rigid-wall solution in the extended DW model in Sec.~\ref{sec:linearstabilitydiagrams}. In both cases, the time variable $\tau = \Omega t$ is discretized with $N_\tau = 1000$ steps per driving period $2\pi$ ($\delta\tau = 2\pi/1000$), ensuring numerical convergence of the limit cycles and Floquet multipliers.

The partial differential equations (PDEs) for the full $u$--$\phi$ model [Eqs.~\eqref{eq:uphi}] were solved using a pseudo-spectral semi-implicit Crank--Nicolson scheme, following the method outlined by Pusiol et al.~\cite{pusiol2026deterministic}. By defining the complex variable~\cite{skaugen2022}
\begin{equation}
    z(x,t)=u(x,t)-i\phi(x,t)\, ,
    \label{eq:z_def}
\end{equation}
the $u$--$\phi$ model is equivalent to the single complex PDE
\begin{equation}
    \dot z = \frac{\alpha-i}{2}\left[\alpha h(t) - i \sin\left(2\operatorname{Im}(z)\right)+z''\right]\, .
    \label{eq:z_eq}
\end{equation}

We separate the right-hand side of Eq.~\eqref{eq:z_eq} into a linear spatial-diffusion term and a nonlinear local term:
\begin{align}
    \mathcal L(z) &= \frac{\alpha-i}{2}z'' \, , \label{eq:lin_term}\\
    \mathcal{N}(z)&=\frac{\alpha-i}{2}\left[\alpha h(t)-i\sin\left(2\operatorname{Im}(z)\right)\right]\, . \label{eq:nonlin_term}
\end{align}
The linear term is stepped according to the Crank--Nicolson method, while the nonlinear term is evaluated explicitly using an Euler step:
\begin{equation}
\frac{z^{n+1}-z^{n}}{\delta t}=\frac{1}{2} \mathcal L\left(z^{n+1}+z^{n}\right)+\mathcal N(z^n)\, .
    \label{eq:stepping}
\end{equation}
Transforming to spatial Fourier space, where $\hat{\mathcal{L}}_k = -\frac{\alpha-i}{2}k^2$, we obtain an explicit algebraic update formula for the $k$-th Fourier mode:
\begin{equation}
    z^{n+1}_k=\frac{\left(1+\frac{\delta t}{2}\hat{\mathcal L}_k\right)z_k^n+\delta t \widehat{\mathcal N(z^n)}_k}{1-\frac{\delta t}{2} \hat{\mathcal L}_k} \, .
    \label{eq:fourier_update}
\end{equation}
At each time step, the nonlinear term $\mathcal{N}(z^n)$ is computed locally in real space and then transformed to Fourier space via the Fast Fourier Transform (FFT). The updated field $z_k^{n+1}$ is then transformed back to real space.

The 1D spatial domain of length $L$ (with periodic boundary conditions) is discretized into $N$ grid points with a uniform grid spacing $\delta x = L/N = 0.5\,L_0$ ($N = 2 L/L_0$), which accurately resolves the domain-wall profile and all relevant flexural modes up to the continuum cutoff. The simulation time step was fixed to $\delta t = 2\pi/(320\,\Omega) \approx 0.019635$, corresponding to 320 time steps per driving period $T_\text{drive} = 2\pi/\Omega$. Numerical stability and convergence were systematically verified against both spatial grid refinement (halving $\delta x$ to $0.25\,L_0$) and time-step refinement (halving $\delta t$ down to $2\pi/(1280\,\Omega)$), confirming that the computed mean velocities, roughness, and phase boundaries are insensitive to the discretization.

Initial conditions consisted of flat domain walls with small random transverse perturbations of amplitude $\epsilon = 5\times 10^{-4}$ in both $u$ and $\phi$.

We implemented this solver in C++ (with OpenMP multithreading for CPU execution) and CUDA C++ (using cuFFT with double precision for GPU execution), developing a unified backend-agnostic architecture. The GPU backend provides substantial speedups for large systems, making simulations in the thermodynamic limit up to $L/L_0 = 1024$ (and up to $L/L_0 \approx 2.6\times 10^5$ for finite-size scaling) computationally efficient. The complete source code for the simulator is freely available in Ref.~\onlinecite{github_luz}.

\section{Finite Size Effects}
\label{ap:finitesizephasediagram}

\begin{figure}[tb]
    \centering
    \includegraphics[width=\linewidth]{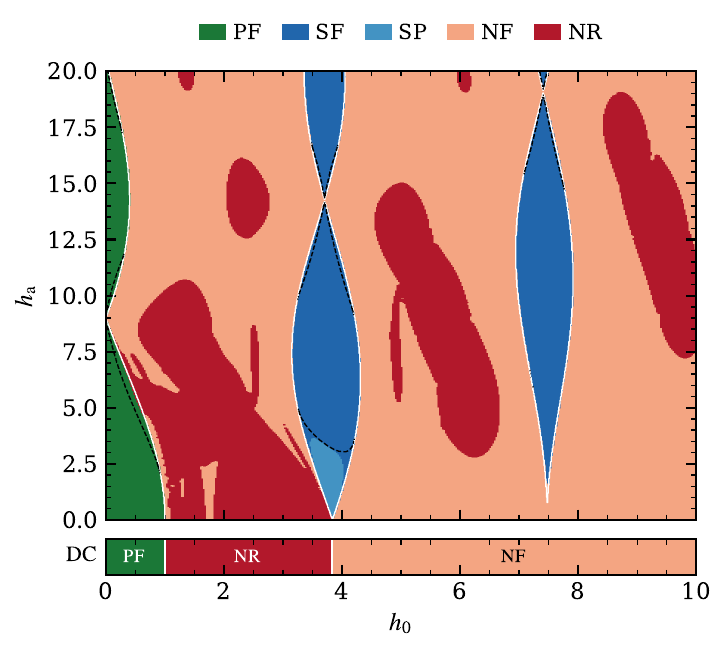}
    \caption{Dynamical phase diagram for a small extended domain wall of size $L/L_0 = 8$, driven by combined DC and AC magnetic fields with $\alpha=0.27$ and $\Omega=1$. The five dynamical regimes are: pinned flat (PF), synchronized flat (SF), synchronized periodic (SP), non-synchronized flat (NF), and non-synchronized rough (NR). Continuous white and dashed black curves indicate the rigid-wall synchrony and linear-stability boundaries, respectively. The horizontal strip at the bottom corresponds to purely DC driving ($h_\mathrm{a} = 0$), showing the non-precessional regime (PF, $0 \le h_0 \le 1$), the precessional rough regime (NR, $1 < h_0 < h_c$), and the precessional flat regime (NF, $h_0 \ge h_c$), where $h_c \approx 3.8$ for $L/L_0=8$ due to the discrete mode spectrum.}
    \label{fig:phasediagram_smallL}
\end{figure}

In Figure~\ref{fig:phasediagram_smallL}, we show the dynamical phase diagram for a relatively small system, $L/L_0=8$. The differences with the large-system phase diagram (Figure~\ref{fig:fases}) highlight the pronounced role of finite-size effects. In particular, for the small system, the discrete set of allowed spatial modes $\kappa_n = 2\pi n / (L/L_0) \approx 0.785 n$ limits the availability of modes that can fall within the linear instability band of the precessing DW. Under purely DC driving ($h_\mathrm{a} = 0$), shown in the horizontal strip below the diagram, the first flexural mode $\kappa_1 \approx 0.785$ is only unstable within a restricted window $1 < h_0 < h_c$, where $h_c \approx 3.8$. For $h_0 > h_c$, all available discrete modes become linearly stable, so the domain wall precesses as a flat interface (NF phase), in sharp contrast with the thermodynamic limit ($L\to\infty$), where the dense spectrum of long-wavelength modes sustains the chaotic NR regime up to $h_c \approx 7$. Similarly, under AC driving, the SF and NF phases dominate a broader region of the phase diagram in the small system, while the chaotic NR regime is significantly reduced compared to large systems. 

\section{Second-order perturbation theory for rigid DW synchrony}
\label{ap:segundo_orden}

We can continue the expansion of Eq.~\eqref{eq:expansion2} to second order in $\epsilon$. To do this, we expand $\epsilon\sin(\psi_1) \approx \epsilon \sin\left(\psi^{(0)}\right) + \epsilon^2 \psi^{(1)} \cos\left(\psi^{(0)}\right)$, substitute into Eq.~\eqref{eq:walker_pert} and collect the terms of order $\epsilon^2$. We obtain:
\begin{equation}
\dot{\psi}^{(2)} = \tilde h_0^{(2)} - \psi^{(1)} \cos\left(\psi^{(0)}\right).
    \label{eq:dot_psi_2}    
\end{equation}

To find $\psi^{(1)}$, we integrate Eq.~\eqref{eq:dot_psi_1} using the Jacobi--Anger expansion for $\sin\left(\psi^{(0)}\right)$ and substituting $\tilde h_0^{(1)}$ from Eq.~\eqref{eq:firstorder}. This yields:
\begin{equation}
    \psi^{(1)}(\tau) = \sum_{m \neq -n} \frac{J_m(\tilde h_\text{a})}{n+m} \cos[(n+m)\tau + \psi_\text{c}].
\end{equation}

In order for the second-order solution to remain bounded and $2\pi$-periodic, we require again that the time average of Eq.~\eqref{eq:dot_psi_2} vanishes: $\tilde h_0^{(2)} = \left\langle \psi^{(1)} \cos\left(\psi^{(0)}\right) \right\rangle_\tau$. To compute this average, we expand $\cos\left(\psi^{(0)}\right)$ analogously to the sine term:
\begin{equation}
    \cos\left(\psi^{(0)}\right) = \sum_{k=-\infty}^{\infty} J_k(\tilde h_\text{a}) \cos[(n+k)\tau + \psi_\text{c}].
\end{equation}

In the product $\psi^{(1)} \cos\left(\psi^{(0)}\right)$, the time average of terms of the form $\langle \cos(p\tau+\psi_\text{c}) \cos(q\tau+\psi_\text{c}) \rangle_\tau$ is non-zero only when $p=\pm q$. Filtering these resonant terms leads to two contributions to the second-order correction:
\begin{equation}
    \tilde h_0^{(2)} = A + B \cos(2\psi_\text{c}),
    \label{eq:secondorder_h0}
\end{equation}
where, using the identity $J_{-v}(x) = (-1)^v J_v(x)$, we define the summation constants $A$ and $B$ as:
\begin{align}
    A &= \frac{1}{2}\sum_{m\neq -n} \frac{J_m^2(\tilde h_\text{a})}{n+m}, \\
    B &= \frac{1}{2}\sum_{m\neq -n} (-1)^m \frac{J_m(\tilde h_\text{a})J_{2n+m}(\tilde h_\text{a})}{n+m}.
\end{align}

However, shifting the index in our expression for $B$ to $r=n+m$, we can rewrite to obtain:
\begin{equation}
    B=\frac{1}{2}\sum_{r\neq 0} \frac{J_{r-n}(\tilde h_\text{a})J_{-r-n}(\tilde h_\text{a})}{r}=0,
\end{equation}
since the sum terms are odd in $r$. This simplifies Eq.~\eqref{eq:secondorder_h0} to $\tilde{h}_0^{(2)}=A$. Also note that $A=0$ if $n=0$, for similar reasons. The second-order corrections vanish in the zeroth synchrony step.

Substituting into Eq.~\eqref{eq:secondorder_h0}, the DC field in the synchrony condition becomes:
\begin{equation}
    \tilde h_0 \approx n + \epsilon^2A + \epsilon(-1)^n J_n(\tilde h_\text{a})\sin(\psi_c),
\end{equation}
or, in the original physical variables,
\begin{equation}
    h_0 \approx \frac{n\Omega}{\alpha} + \frac{\alpha}{\Omega}A + (-1)^n J_n\left(\frac{\alpha}{\Omega} h_\text{a}\right)\sin(\psi_c).
    \label{eq:h0_2order}
\end{equation}
This is the same as the first-order case, but with a rigid, field-dependent offset in the center of the synchronization step $\frac{\alpha}{\Omega}A$. Consequently, the only second-order effect is a rigid shift in the synchronization steps, depending only on $\alpha h_\text {a}/\Omega$, while maintaining their widths.

In the limit $h_\text{a}=0$, the only term contributing to $A$ is the one proportional to $J_0$. For $n\neq0$, $A=\frac{1}{2n}$: the shift is towards higher absolute DC fields and decreases with the absolute value of the step index. In the limit of $h_\text{a}\to\infty$, $A\to 0$ due to the decay of the Bessel functions, implying that the second-order corrections are most significant for small AC fields.

Figure~\ref{fig:ej_walker} compares the first- and second-order approximations for the Shapiro step locations.

\bibliography{refs}

@article{Schryer1974,
  author    = {N. L. Schryer and L. R. Walker},
  title     = {The motion of 180\textdegree{} domain walls in uniform dc magnetic fields},
  journal   = {Journal of Applied Physics},
  volume    = {45},
  number    = {12},
  pages     = {5406--5421},
  year      = {1974},
  doi       = {10.1063/1.1663252},
  url       = {https://doi.org/10.1063/1.1663252},
  publisher = {American Institute of Physics}
}

@inproceedings{Slonczewski1972,
  author    = {J. C. Slonczewski},
  title     = {Dynamics of Magnetic Domain Walls},
  booktitle = {AIP Conference Proceedings},
  volume    = {5},
  pages     = {170--174},
  year      = {1972},
  doi       = {10.1063/1.2946580},
  url       = {https://doi.org/10.1063/1.2946580},
  publisher = {American Institute of Physics}
}

@article{skaugen2022,
  title = {Depinning Exponents of Thin Film Domain Walls Depend on Disorder Strength},
  author = {Skaugen, Audun and Laurson, Lasse},
  journal = {Phys. Rev. Lett.},
  volume = {128},
  issue = {9},
  pages = {097202},
  numpages = {5},
  year = {2022},
  month = {Mar},
  publisher = {American Physical Society},
  doi = {10.1103/PhysRevLett.128.097202},
  url = {https://link.aps.org/doi/10.1103/PhysRevLett.128.097202}
}

@article{ferre2013,
title = {Universal magnetic domain wall dynamics in the presence of weak disorder},
journal = {Comptes Rendus Physique},
volume = {14},
number = {8},
pages = {651-666},
year = {2013},
note = {Disordered systems / Systèmes désordonnés},
issn = {1631-0705},
doi = {https://doi.org/10.1016/j.crhy.2013.08.001},
url = {https://www.sciencedirect.com/science/article/pii/S1631070513001291},
author = {Jacques Ferré and Peter J. Metaxas and Alexandra Mougin and Jean-Pierre Jamet and Jon Gorchon and Vincent Jeudy}
}

@article{Parkin2008,
  title = {Magnetic Domain-Wall Racetrack Memory},
  author = {Parkin, Stuart S. P. and Hayashi, Masamitsu and Thomas, Luc},
  journal = {Science},
  volume = {320},
  number = {5873},
  pages = {190--194},
  year = {2008},
  doi = {10.1126/science.1145799},
  url = {https://www.science.org/doi/10.1126/science.1145799}
}

@article{ferrero2021creep,
  title={Creep motion of elastic interfaces driven in a disordered landscape},
  author={Ferrero, Ezequiel E and Foini, Laura and Giamarchi, Thierry and Kolton, Alejandro B and Rosso, Alberto},
  journal={Annual Review of Condensed Matter Physics},
  volume={12},
  number={1},
  pages={111--134},
  year={2021},
  publisher={Annual Reviews}
}

@article{gourdon2013domain,
  title = {Domain-wall flexing instability and propagation in thin ferromagnetic films},
  author = {Gourdon, C. and Thevenard, L. and Haghgoo, S. and C\ifmmode \bar{e}\else \={e}\fi{}bers, A.},
  journal = {Phys. Rev. B},
  volume = {88},
  issue = {1},
  pages = {014428},
  numpages = {9},
  year = {2013},
  month = {Jul},
  publisher = {American Physical Society},
  doi = {10.1103/PhysRevB.88.014428},
  url = {https://link.aps.org/doi/10.1103/PhysRevB.88.014428}
}

@article{thevenard2011domain,
  title = {Domain wall propagation in ferromagnetic semiconductors: Beyond the one-dimensional model},
  author = {Thevenard, L. and Gourdon, C. and Haghgoo, S. and Adam, J-P. and von Bardeleben, H. J. and Lema\^{\i}tre, A. and Schoch, W. and Thiaville, A.},
  journal = {Phys. Rev. B},
  volume = {83},
  issue = {24},
  pages = {245211},
  numpages = {9},
  year = {2011},
  month = {Jun},
  publisher = {American Physical Society},
  doi = {10.1103/PhysRevB.83.245211},
  url = {https://link.aps.org/doi/10.1103/PhysRevB.83.245211}
}

@article{Kim2024,
  title = {Enhancement of magnetic domain wall velocity via resonant dissipation of standing wave modes of domain wall structure with perpendicular magnetic anisotropy},
  author = {Kim, Ganghwi and Jung, Dae-Han and Han, Hee-Sung and Kang, Myeonghwan and Jeong, Suyeong and Park, Younggun and Im, Mi-Young and Lee, Ki-Suk},
  journal = {Phys. Rev. B},
  volume = {110},
  issue = {21},
  pages = {214407},
  numpages = {10},
  year = {2024},
  month = {Dec},
  publisher = {American Physical Society},
  doi = {10.1103/PhysRevB.110.214407},
  url = {https://link.aps.org/doi/10.1103/PhysRevB.110.214407}
}

@article{skaugen2019,
  title = {Analytical computation of the demagnetizing energy of thin-film domain walls},
  author = {Skaugen, Audun and Murray, Peyton and Laurson, Lasse},
  journal = {Phys. Rev. B},
  volume = {100},
  issue = {9},
  pages = {094440},
  numpages = {10},
  year = {2019},
  month = {Sep},
  publisher = {American Physical Society},
  doi = {10.1103/PhysRevB.100.094440},
  url = {https://link.aps.org/doi/10.1103/PhysRevB.100.094440}
}

@book{Strogatz2015,
  author    = {Steven H. Strogatz},
  title     = {Nonlinear Dynamics and Chaos: With Applications to Physics, Biology, Chemistry, and Engineering},
  edition   = {2},
  publisher = {Westview Press},
  year      = {2015},
  isbn      = {978-0813349107}
}

@article{pusiol2026deterministic,
  title={Deterministic roughening in the dc-driven precessional regime of domain walls},
  author={Pusiol, EF and Lecomte, Vivien and Bustingorry, S and Kolton, AB},
  journal={Physical Review B},
  volume={113},
  number={6},
  pages={064411},
  year={2026},
  publisher={APS}
}

@book{press2007numerical,
  title = {Numerical Recipes 3rd Edition: The Art of Scientific Computing},
  author = {Press, William H. and Teukolsky, Saul A. and Vetterling, William T. and Flannery, Brian P.},
  year = {2007},
  edition = {3},
  publisher = {Cambridge University Press},
  address = {New York, NY, USA},
  isbn = {978-0-521-88068-8}
}

@misc{github_luz,
  author = {{Luz Pérez Herrera}},
  title = {{DW-uphi-simulator}},
  year = {2026},
  publisher = {GitHub},
  journal = {GitHub repository},
  howpublished = {\url{https://github.com/lightph/dw-uphi-simulator}},
}

@article{koltonmodelocking,
  title = {Mode Locking in ac-Driven Vortex Lattices with Random Pinning},
  author = {Kolton, Alejandro B. and Dom\'{\i}nguez, Daniel and Gr\o{}nbech-Jensen, Niels},
  journal = {Phys. Rev. Lett.},
  volume = {86},
  issue = {18},
  pages = {4112--4115},
  numpages = {0},
  year = {2001},
  month = {Apr},
  publisher = {American Physical Society},
  doi = {10.1103/PhysRevLett.86.4112},
  url = {https://link.aps.org/doi/10.1103/PhysRevLett.86.4112}
}

@article{koltonmodelocking2,
  title = {Mode locking in driven vortex lattices with transverse ac drive and random pinning},
  author = {Kolton, Alejandro B. and Dom\'{\i}nguez, Daniel and Gr\o{}nbech-Jensen, Niels},
  journal = {Phys. Rev. B},
  volume = {65},
  issue = {18},
  pages = {184508},
  numpages = {5},
  year = {2002},
  month = {Apr},
  publisher = {American Physical Society},
  doi = {10.1103/PhysRevB.65.184508},
  url = {https://link.aps.org/doi/10.1103/PhysRevB.65.184508}
}

@article{okumasuppresion,
  title = {Suppression of broadband noise at mode locking in driven vortex matter},
  author = {Okuma, S. and Inoue, J. and Kokubo, N.},
  journal = {Phys. Rev. B},
  volume = {76},
  issue = {17},
  pages = {172503},
  numpages = {4},
  year = {2007},
  month = {Nov},
  publisher = {American Physical Society},
  doi = {10.1103/PhysRevB.76.172503},
  url = {https://link.aps.org/doi/10.1103/PhysRevB.76.172503}
}

@article{shapiro1963josephson,
  title={Josephson currents in superconducting tunneling: The effect of microwaves and other observations},
  author={Shapiro, Sidney},
  journal={Physical Review Letters},
  volume={11},
  number={2},
  pages={80--82},
  year={1963},
  publisher={APS},
  doi={10.1103/PhysRevLett.11.80}
}

@article{berger1991prediction,
  title={Prediction of Shapiro steps in the dc voltage across a magnetic domain wall traversed by a dc current and exposed to high-frequency magnetic fields},
  author={Berger, L},
  journal={Journal of applied physics},
  volume={69},
  number={8},
  pages={4683--4685},
  year={1991},
  publisher={American Institute of Physics}
}

@article{Munsu2021,
  title = {Domain-wall motion driven by a rotating field in a ferrimagnet},
  author = {Jin, Munsu and Hong, Ik-Sun and Kim, Duck-Ho and Lee, Kyung-Jin and Kim, Se Kwon},
  journal = {Phys. Rev. B},
  volume = {104},
  issue = {18},
  pages = {184431},
  numpages = {7},
  year = {2021},
  month = {Nov},
  publisher = {American Physical Society},
  doi = {10.1103/PhysRevB.104.184431},
  url = {https://link.aps.org/doi/10.1103/PhysRevB.104.184431}
}

@misc{kim2019,
      title={Magnetic Domain Wall Motion due to AC Bias-Driven Resonances}, 
      author={Duck-Ho Kim and Dong-Hyun Kim and Dae-Yun Kim and Sug-Bong Choe and Teruo Ono and Kyung-Jin Lee and Se Kwon Kim},
      year={2019},
      eprint={1908.01264},
      archivePrefix={arXiv},
      primaryClass={cond-mat.mes-hall},
      url={https://arxiv.org/abs/1908.01264}, 
}

@article{Moon_2017,
doi = {10.1088/1361-6463/aa5d35},
url = {https://doi.org/10.1088/1361-6463/aa5d35},
year = {2017},
month = {feb},
publisher = {IOP Publishing},
volume = {50},
number = {12},
pages = {125003},
author = {Moon, Kyoung-Woong and Kim, Duck-Ho and Kim, Changsoo and Kim, Dae-Yun and Choe, Sug-Bong and Hwang, Chanyong},
title = {Domain wall motion driven by an oscillating magnetic field},
journal = {Journal of Physics D: Applied Physics}
}

@book{Pikovsky_Rosenblum_Kurths_2001, place={Cambridge}, series={Cambridge Nonlinear Science Series}, title={Synchronization: A Universal Concept in Nonlinear Sciences}, publisher={Cambridge University Press}, author={Pikovsky, Arkady and Rosenblum, Michael and Kurths, Jürgen}, year={2001}, collection={Cambridge Nonlinear Science Series}}

@article{adler1946study,
  title={A study of locking phenomena in oscillators},
  author={Adler, Robert},
  journal={Proceedings of the IRE},
  volume={34},
  number={6},
  pages={351--357},
  year={1946},
  publisher={IEEE}
}

\end{document}